\documentclass[12pt]{article}

\usepackage[numbers]{natbib}
\usepackage{amsmath}
\usepackage{orcidlink}
\usepackage{booktabs}

\usepackage{authblk}      

\usepackage{tikz}
\usetikzlibrary{shapes.geometric, arrows, positioning}

\usepackage{graphicx}
\usepackage{geometry}
\usepackage{ragged2e}     
\AtBeginDocument{\justifying}   

\begin{document}
\graphicspath{{figs/}}


\title{Oval-shaped resonance distortion as a signature of quasiparticle heating effect in a niobium superconducting resonator}

\author{
Zhenyuan Sun$^{\dagger,1,2}$\orcidlink{0000-0002-1370-2074},
Genting Dai$^1$,
Xiao Geng$^1$,
Liangliang Yang$^1$,
Mingjun Cheng$^1$,
Qing Yu$^3$,
Jinlin Chang$^1$,
Yi Yang$^1$,
Linpan Jiang$^1$,
Jianshe Liu$^1$,
and Wei Chen$^*$
}

\affil[1]{School of Integrated Circuits, Tsinghua University, 100084, Beijing, China}
\affil[2]{Cavendish Laboratory, University of Cambridge, CB3 0HE, Cambridge, UK}
\affil[3]{Department of Astronomy, Tsinghua University, 100084, Beijing, China}
\affil[$^{\dagger}$]{zs311@cantab.ac.uk, $^*$weichen@mail.tsinghua.edu.cn}
\affil[ ]{2026.07.20}

\date{}   


\maketitle

\begin{abstract}
We investigate the nonlinear behavior of a superconducting microwave resonator subjected to a dissipative mechanism where the associated quality factor (Q factor) decreases with increasing dissipated power, leading to a dissipative feedback effect. By modifying the Rothwarf-Taylor equations, we establish a macroscopic quasiparticle heating (QPH) model that directly links the quality factor to the microwave readout power. The key finding is the identification of a distinctive oval-shaped distortion in the resonance circle in the complex plane. This distortion serves as a practical experimental signature for identifying the readout power regime in which QPH dominates the loss, under conditions where other nonlinear mechanisms are sufficiently weak. To validate the model, we design and fabricate a niobium (Nb) half-wavelength coplanar waveguide (CPW) resonator and conduct systematic bath temperature and readout power sweeps. The model provides a well fit to the observed oval-shaped resonance circle distortion across a wide range of operating conditions, confirming the QPH mechanism as the primary source of the dissipative non-linearity in the parameter space investigated.
\end{abstract}


\section{Introduction}\label{sec_introduction}
Superconducting resonators have been widely adopted in both fundamental physics research and practical applications, such as kinetic inductance detectors (KIDs), solid-state qubits and readout circuits for cryogenic devices \cite{zmuidzinas2012superconducting, blais2021circuit, irwin2004microwave}.  These resonators are typically driven by a readout tone, and increasing the readout power often provides significant advantages. For instance, in the case of KIDs, higher power helps suppress noise originating from two-level systems (TLS) and readout electronics \cite{gao2008semiempirical, mchugh2012readout}. The practical limit is generally set by the onset of nonlinear behaviour,  where resonator properties, such as resonance frequency and Q factor, begin to deviate from their linear behaviour. Therefore, a detailed understanding of the physical mechanisms underlying the nonlinear dynamics, the corresponding power thresholds, and the resulting degradation or enhancement \cite{swenson2013operation} of device performance in the nonlinear regime is of significant importance.  

Reactive non-linearity, whereby the readout power affects the resonant frequency, is a well-documented phenomenon in superconducting microresonators, commonly attributed to intrinsic non-linearity in the kinetic inductance \cite{swenson2013operation, PhysRevLett.117.047002, pippard1950field}, quasiparticle heating \cite{de2010readout, goldie2012non,thompson2013dynamical}, and superconducting weak links \cite{abdo2005unexpected}. Swenson’s model\cite{swenson2013operation} , which incorporates Duffing oscillator dynamics, offers a straightforward analytical framework and has become a widely used fitting tool in the KID community.

By comparison, dissipative nonlinear behaviour, whereby the readout power affects the quality factor, warrants further investigation, as it plays a critical role in limiting the sensitivity of superconducting devices. Several mechanisms have been described in the literature, including two-level systems in dielectric parts of the resonator \cite{gao2008physics, pappas2011two} and quasiparticle generation from indirect pair-breaking by the readout signal \cite{goldie2012non}, known as the quasiparticle heating process, which account for measured quality factor data over limited power ranges. In the context of resonator dynamics, a key observation is that the consequential macroscopic behaviour can be described by a reduced model where the quasiparticles are ascribed an effective temperature above their physical temperature. The power dissipated by the readout signal effectively heats the quasiparticles \cite{de2010readout}, and an equilibrium state is formed when the heating power is balanced by the cooling power flow to the phonons \cite{goldie2012non, guruswamy2015nonequilibrium}. This electrothermal model has been used to account for both large-signal \cite{de2010readout, thompson2013dynamical, fischer2024nonequilibrium}  and small-signal \cite{thomas2015electrothermal, catelani2019non, fischer2023nonequilibrium} device behaviour. Little seems to have been published on the resonance distortion in presence of dissipative nonlinearities, which is important both for fitting and as a syndrome that reveals the action of a particular mechanism \cite{thomas2020nonlinear}.

In this paper, we address the issue of resonance-curve distortion in the presence of dissipative nonlinearities by investigating quasiparticle heating as a key mechanism in a Nb CPW half-wavelength resonator. Specifically, we consider the case where the quality factor decreases with increasing readout power (equivalently, $Q$ factor is inversely proportional to the quasiparticle density), so that as more power is dissipated, the device becomes lossier. In Sec.\ref{sec_2_QPH_model}, we show that this mechanism produces both a characteristic oval-shaped distortion of the resonance circle in the complex plane and a distinct power-law dependence of the quality factor on readout power. Experimental observations of this behaviour in our fabricated device are presented in Sec.\ref{sec_3_measurement}, demonstrating that the predicted distortion provides a practical diagnostic signature of the QPH-dominated regime over a wide range of operating conditions, while also noting that additional mechanisms may become relevant at the highest power levels. 

\section{Quasiparticle heating model}\label{sec_2_QPH_model}
For a two-port peak-type resonator,  the forward transmission scattering parameter is \cite{thomas2020nonlinear, pozar2011microwave, khalil2012analysis} 
\begin{equation}\label{eqn:S1_s_parameter_halfwave_resonator}
S_{21} = \frac{Q_r}{Q_c} \frac{1}{1 + 2 i Q_r x},
\end{equation}
with the total quality factor $Q_r$ defined by $Q_r^{-1} = Q_i^{-1} + Q_c^{-1}$,  $Q_c$ the coupling quality factor, $Q_i$ the internal quality factor and $x$ the fractional frequency detuning defined by $x = (f -f_r) / f_r$ where $f$ is the frequency of the readout signal and $f_r$ is the resonance frequency. When the readout power $P_r$ is applied, the power dissipated into the resonator is given by
\begin{equation}\label{eqn:S1_dissipated_power}
P_{diss} = \frac{2Q_r}{Q_c}\,\frac{1}{1 + (2Q_r x)^2}\,\frac{Q_r}{Q_i}\,P_r,
\end{equation}
The total internal loss is partitioned as $Q_i^{-1} = Q_{qp}^{-1} + Q_{other}^{-1}$, where $Q_{qp}$ accounts for dissipation due to quasiparticles and $Q_{other}$ subsumes all other loss channels. The power flowing into the quasiparticle system is therefore $P_{qp} = P_{diss}\,Q_i/Q_{qp}$. Given that Mattis-Bardeen theory predicts $Q_{qp}$ to be inversely proportional to $n_{qp}$ as the operating bath temperature $T_b$ is well below the critical transition temperature $T_c$ \cite{gao2008physics, mattis1958theory, noguchi2016effect, noguchi2018analysis}, this proportionality motivates the introduction of a scaling factor $n_*$, including all the effects of temperature, frequency, kinetic inductance and resonator geometry.  It is defined as the critical quasiparticle density at which $Q_{qp} = Q_c$, i.e. $Q_{qp} = n_* Q_c / n_{qp}$. 

The theoretical framework for describing the quasiparticle heating process in superconducting resonators is established by introducing a quasiparticle generation-rate term $\Gamma_r$ \cite{thomas2020nonlinear, sun2026quasiparticle}, builds upon the Rothwarf--Taylor equations \cite{rothwarf1967measurement}.  
\begin{equation}\label{eqn:S1_rothwarf_taylor_equation_1}
\frac{\partial n_{qp}}{\partial t} = \frac{2}{\tau_{pb}}n_\omega - Rn^2_{qp},
\end{equation}
\begin{equation}\label{eqn:S1_rothwarf_taylor_equation_2}
\frac{\partial n_{\omega}}{\partial t} = -\frac{1}{\tau_{pb}}n_\omega + \frac{R}{2} n^2_{qp} - \frac{1}{\tau_{l}} [n_\omega - n_{\omega, th}] + \Gamma_r.
\end{equation}
Here $n_{\mathrm{qp}}$ is the quasiparticle number density, $n_{\omega}$ is the number density of phonons with energy exceeding the pair-breaking threshold occupying the same active volume $V$ of the resonator,  and $n_{\omega,\mathrm{th}}$ is the thermal-equilibrium value of $n_{\omega}$ in the absence of external driving ($\Gamma_r = 0$). The characteristic time scales are the pair-breaking time $\tau_{\mathrm{pb}}$, the phonon lifetime $\tau_l$ in the absence of interactions with the quasiparticle system, and $R$ is the bimolecular quasiparticle recombination rate. 

We assume steady-state operation ($\partial n_{\mathrm{qp}}/\partial t = 0$, $\partial n_{\omega}/\partial t = 0$) and readout photons themselves are far below the pair-breaking frequency threshold, so that direct photon-induced pair breaking is negligible. Eliminating $n_{\omega}$ by substituting Eq.\ref{eqn:S1_rothwarf_taylor_equation_2} into Eq.\ref{eqn:S1_rothwarf_taylor_equation_1} yields the compact governing equation
\begin{equation}\label{eqn:S1_governing_equation_with_forcing_simplification}
R [n_{qp}^2 -  n_{qp,th}^2]= \frac{2\tau_l}{\tau_{pb}} \Gamma_r\,.
\end{equation}
Recognizing that in the low readout-power limit ($\Gamma_r = 0$) the quasiparticle density in thermal equilibrium must recover its thermal value $n_{\mathrm{qp,th}}$(corresponding quality factor $Q_{qp,th} = n_* Q_c / n_{qp,th}$), we have $R\,n_{\mathrm{qp,th}}^{2} = 2n_{\omega,\mathrm{th}}/\tau_{\mathrm{pb}}$.  Eq.\ref{eqn:S1_governing_equation_with_forcing_simplification} captures the key balance governing the steady-state quasiparticle density: the net recombination rate (left-hand side) is proportional to the phonon-mediated generation rate driven by the readout signal (right-hand side). 

A key modification in the model is the introduction of the generation rate term as $\Gamma_r = \eta P_{\mathrm{qp}}/(V\Delta)$,  at which pair-breaking phonons are generated from the microwave readout signal with power $P_r$ via quasiparticle-phonon elastic scattering \cite{guruswamy2015nonequilibrium}.  Here $\eta$ is a dimensionless generation efficiency parameter, $P_{qp}$ is the power dissipated into quasiparticle system by the readout signal, and $\Delta$ is the superconducting energy gap defined as $\Delta \approx 1.76\,k_B T_c$, where $k_B$ is the Boltzmann constant.  Given Eq.\ref{eqn:S1_dissipated_power}, the generation rate at detuning $x$ is
\begin{equation}\label{eqn:S1_generation_term_at_detuning}
\Gamma_r = \frac{1}{1 + 4 Q_c^2 x^2 (\frac{n_{qp}}{n_*} + \frac{Q_c}{Q_{other}} + 1)^{-2}} \cdot \frac{n_{\mathrm{qp}}/n_*}{\bigl[n_{\mathrm{qp}}/n_* + 1 + Q_c/Q_{\mathrm{other}}\bigr]^{2}}\, \frac{2\eta P_r}{V\Delta}. 
\end{equation}

For a fabricated device with a predetermined $Q_c$,  substituting Eq.\ref{eqn:S1_generation_term_at_detuning} into Eq.\ref{eqn:S1_governing_equation_with_forcing_simplification} yields the governing equation at zero detuning ($x = 0$)
\begin{equation}\label{eqn:S1_governing_equation_normalised_zero_detuning}
\resizebox{0.95\width}{!}{$\begin{aligned}
u^4 &+ 2 a u^3 + (a^2 - u_{th}^2) u^2 - (2 a u_{th}^2 + \gamma) u - a^2 u_{th}^2 = 0\\
&u = \frac{n_{qp}}{n_{\ast}}, \,\, u_{th} = \frac{n_{qp,th}}{n_{\ast}}, \,\, a = 1 + \frac{Q_c}{Q_{other}},\,\, \gamma = \frac{P_r}{P_c},
\end{aligned}$}
\end{equation}
where $P_c = \tau_{pb} R V \Delta n_*^2 / 4\eta \tau_l$ is the power scaling factor.  For any given bath temperature $T_b$ (which fixes $u_{\mathrm{th}}$) and readout power $P_r$, the normalised quasiparticle density $u$ is uniquely determined as the physical (real, positive) root of the quartic Eq.\ref{eqn:S1_governing_equation_normalised_zero_detuning} Once $u$ is known, the quasiparticle quality factor follows from $Q_{\mathrm{qp}} = Q_c/u$, and the internal quality factor from $Q_i^{-1} = Q_{\mathrm{qp}}^{-1} + Q_{\mathrm{other}}^{-1}$.  Note that the detailed analytical treatment of the governing equation can be found in \cite{thomas2020nonlinear}, whereas the present work focuses only on the numerical simulation analysis,  which is fully adequate to capture the nonlinear effects arising from sub-gap microwave readout signals.

\begin{figure}[htbp!]
	\centering
    \begin{tikzpicture}
    	\node[inner sep=0, xshift=0cm, yshift=0cm] (image) at (0,0) {\includegraphics[width=.45\textwidth]{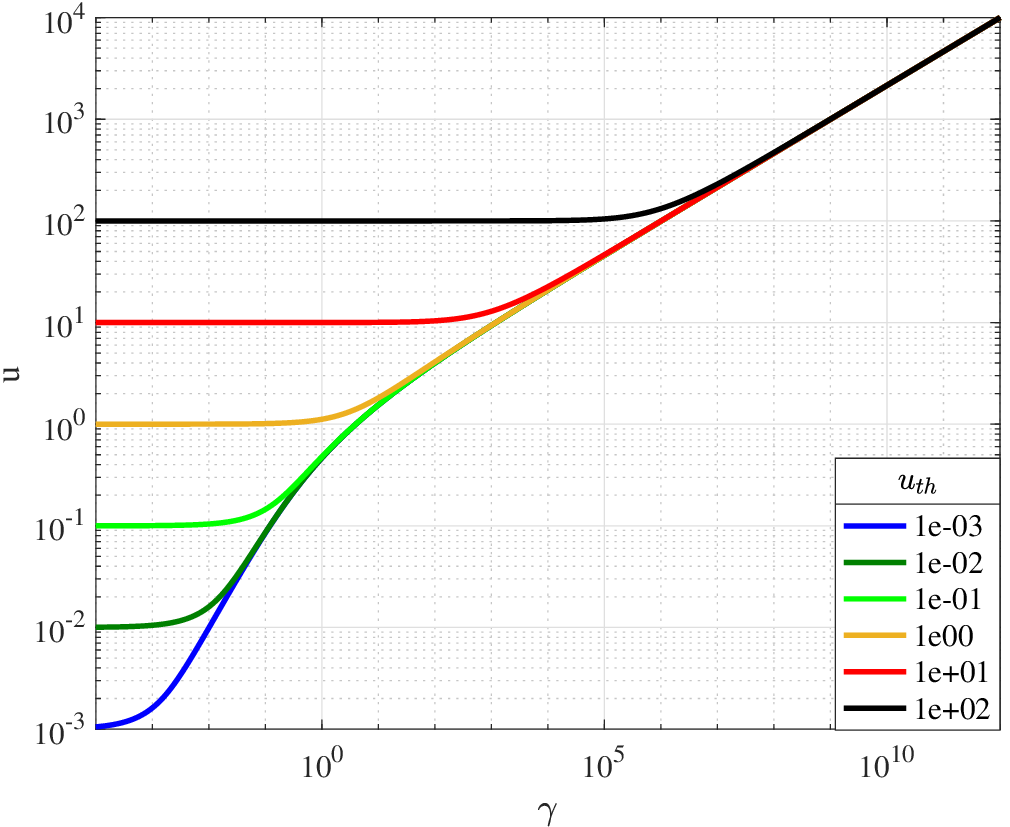}};
    	 \node[inner sep=0, xshift=8cm, yshift=0cm] (image) at (0,0) {\includegraphics[width= .45\textwidth]{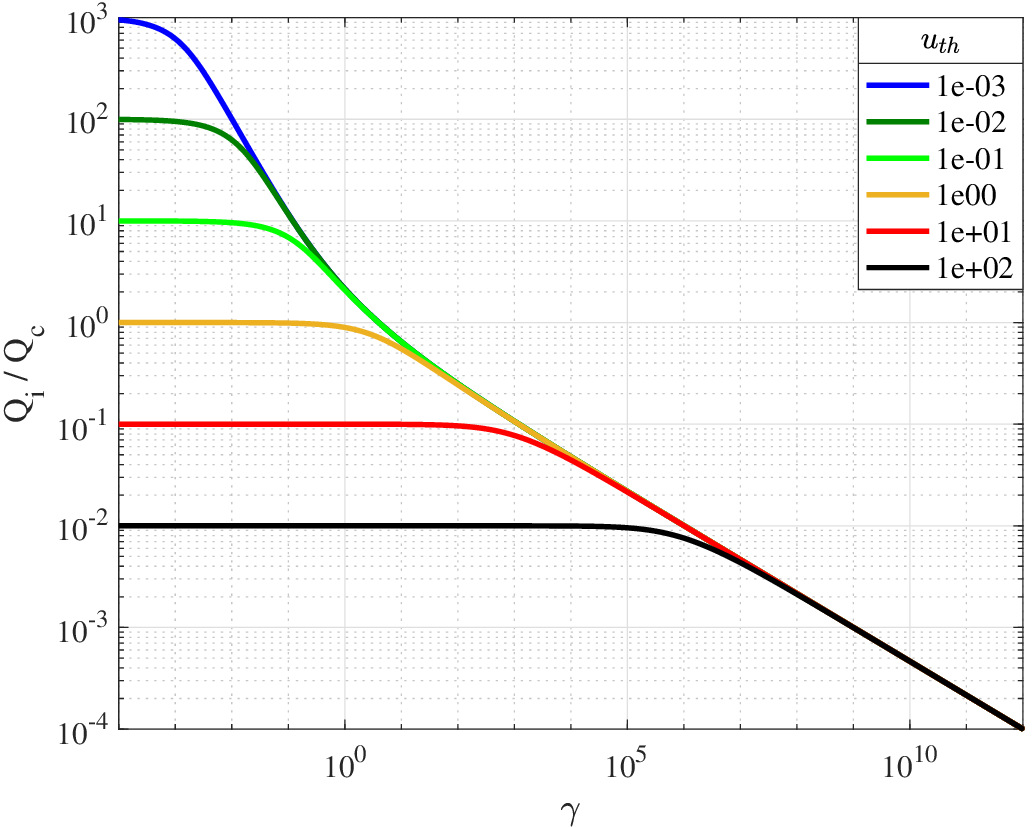}};
    	\node[text=black, scale=1, align=left] at (0, -3) {(a)};
    	\node[text=black, scale=1, align=left] at (8, -3) {(b)};
    	
 	 \end{tikzpicture}	

	\caption[]{\label{fig:S2_n_QiQc_vs_Pr_at_different_T} (a) Normalised quasiparticle density $u = n_{qp} / n_*$ as a function of normalised applied power $\gamma$ for different values of thermal density $u_{th}$. (b) Corresponding quality factor $Q_i / Q_c$. }
	
\end{figure}
Fig.\ref{fig:S2_n_QiQc_vs_Pr_at_different_T} presents numerical solutions of the quartic equation over a wide dynamic range of normalised applied power $\gamma$, for representative values of the thermal density $u_{th}$ (which corresponds to different bath temperatures).  Fig.\ref{fig:S2_n_QiQc_vs_Pr_at_different_T}(a) shows the normalised quasiparticle density $u$,  and Fig.\ref{fig:S2_n_QiQc_vs_Pr_at_different_T}(b) the normalised internal quality factor $Q_i / Q_c$.  Here, $Q_{other} / Q_c = 10^8$ corresponds to the regime where other losses have no influence on the dissipative behaviour of the device.  Several features are noteworthy. At low readout power $\gamma$, the quasiparticle density saturates at the thermal floor $u \rightarrow u_{th}$, and $Q_i / Q_c$ correspondingly exhibits a constant plateau whose level is set by the bath temperature. As readout power increases, the system enters a crossover regime where thermally excited and microwave-induced quasiparticle populations become comparable. At sufficiently high readout power power-induced quasiparticles dominate entirely: all curves collapse onto a universal scaling $u \propto \gamma^{1/3}$ (equivalently $Q_i \propto P_r^{1/3}$),  characteristic of the under-coupled, high-power limit where $Q_i \ll Q_c$. The crossover point from the thermal-dominated to the power-dominated regime shifts to higher readout power as bath temperature increases.

When readout frequency is swept through the resonance, the power coupled into the resonator (and hence the quasiparticle generation rate) varies with the instantaneous detuning.  Considering the non-zero detuning case ($x \neq 0$),  Eq.\ref{eqn:S1_governing_equation_normalised_zero_detuning} is updated into 
\begin{equation}\label{eqn:S1_governing_equation_normalised_non-zero_detuning}
\resizebox{0.85\width}{!}{$u^4 + 2 a u^3 + (a^2 - u_{th}^2 + 4 w^2) u^2 - (2 a u_{th}^2 + \gamma) u - (a^2 + 4 w^2) u_{th}^2 = 0,$}
\end{equation}
where $w = Q_c x$.  At $w = 0$, Eq.\ref{eqn:S1_governing_equation_normalised_non-zero_detuning} reduces to the on-resonance quartic Eq.\ref{eqn:S1_governing_equation_normalised_zero_detuning}.  Solving Eq.\ref{eqn:S1_governing_equation_normalised_non-zero_detuning} point-by-point as the readout frequency is swept through resonance yields the detuning-dependent normalised quasiparticle density $u(w)$, from which the instantaneous internal quality factor $Q_i(w)$ and loaded quality factor $Q_r(w)$ follow. The transmission scattering parameter is then evaluated using the standard resonator response formula Eq.\ref{eqn:S1_s_parameter_halfwave_resonator}, where the on-resonance depth is $|S_{21}(x = 0)| = Q_r/Q_c$, consistent with the usual half-wavelength peak-resonator description.

\begin{figure}[htbp!]
	\centering
    \begin{tikzpicture}
    	\node[inner sep=0, xshift=0cm, yshift=0cm] (image) at (0,0) {\includegraphics[width=.42\textwidth]{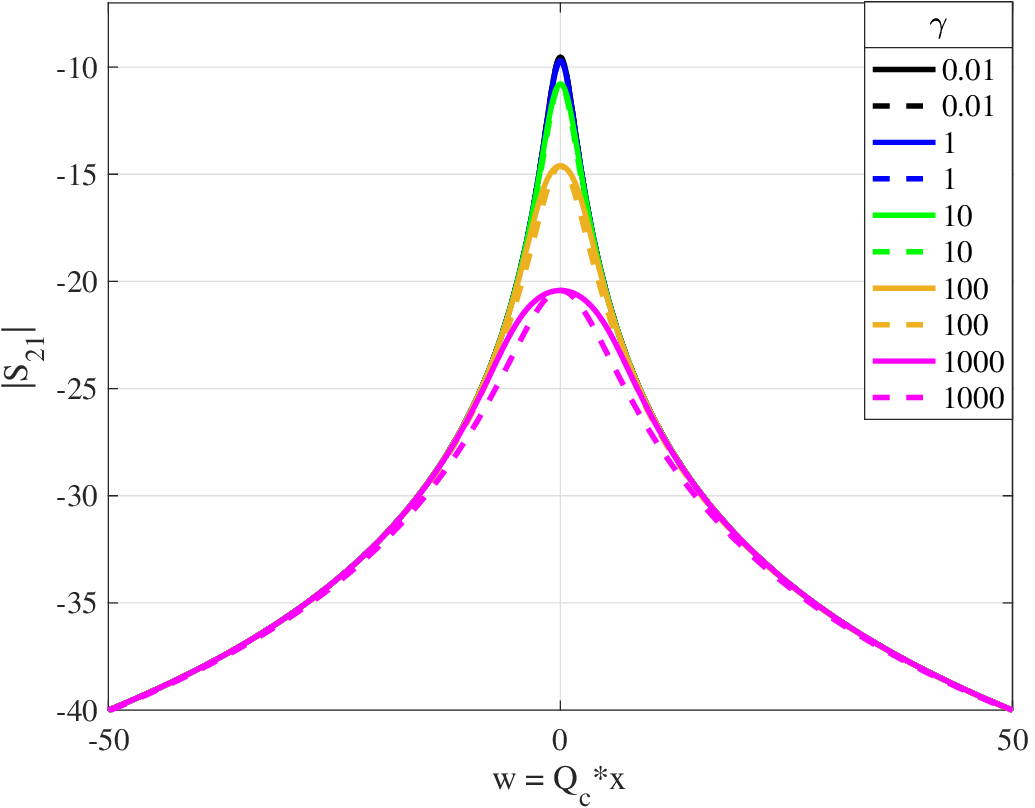}};
    	 \node[inner sep=0, xshift=8cm, yshift=0cm] (image) at (0,0) {\includegraphics[width= .44\textwidth]{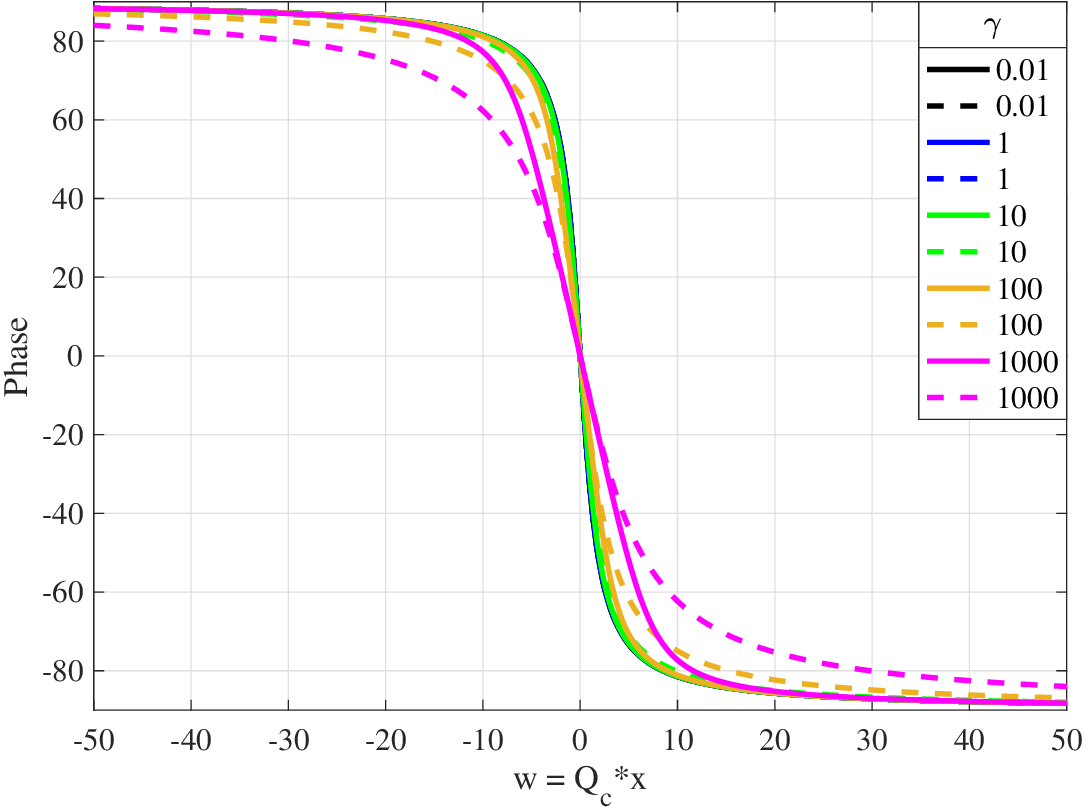}};
    	\node[text=black, scale=1, align=left] at (0, -2.75) {(a)};
    	\node[text=black, scale=1, align=left] at (8, -2.75) {(b)};
    	
 	 \end{tikzpicture}	

	\caption[]{\label{fig:S2_simulation_magnitude_phase_versus_readout_power} Transmission magnitude (a) and phase (b) of $S_{21}$ as a function of frequency at different readout powers.  Solid lines: full QPH model; dashed lines: linear resonance curves calculated by Eq.\ref{eqn:S1_s_parameter_halfwave_resonator} using $Q_c$ and a fixed $Q_r$ obtained from the depth of the matching nonlinear curve as $x = 0$.}
	
\end{figure}
We solved Eq.\ref{eqn:S1_governing_equation_normalised_non-zero_detuning} for the quasiparticle density $n_{qp}$ at an intermediate value of $Q_{qp,th} / Q_c = 0.5$ and then calculated the scattering parameters to simulate how the QPH effect influences the resonance curves, particularly in the high readout power regime where heating is significant. Fig.\ref{fig:S2_simulation_magnitude_phase_versus_readout_power} successfully reproduces the observed decline in resonance depth and the associated distortion, which serve as signatures of the QPH regime. At low readout powers (black and blue curves), sharp resonance curves are obtained. As the applied power increases, the dashed lines (linear reference) rise much faster in magnitude than the solid lines (full QPH model), consistent with an increase in $Q_r$ in the full model when the dissipated power falls off‑resonance. The distortion is more pronounced in the phase response Fig.\ref{fig:S2_simulation_magnitude_phase_versus_readout_power}(b) than in the magnitude response  Fig.\ref{fig:S2_simulation_magnitude_phase_versus_readout_power}(a), reflecting the sensitivity of the phase to changes in the internal quality factor near resonance.

\begin{figure}[htbp!]
	\centering
    \begin{tikzpicture}
    	\node[inner sep=0, xshift=0cm, yshift=0cm] (image) at (0,0) {\includegraphics[width=.425\textwidth]{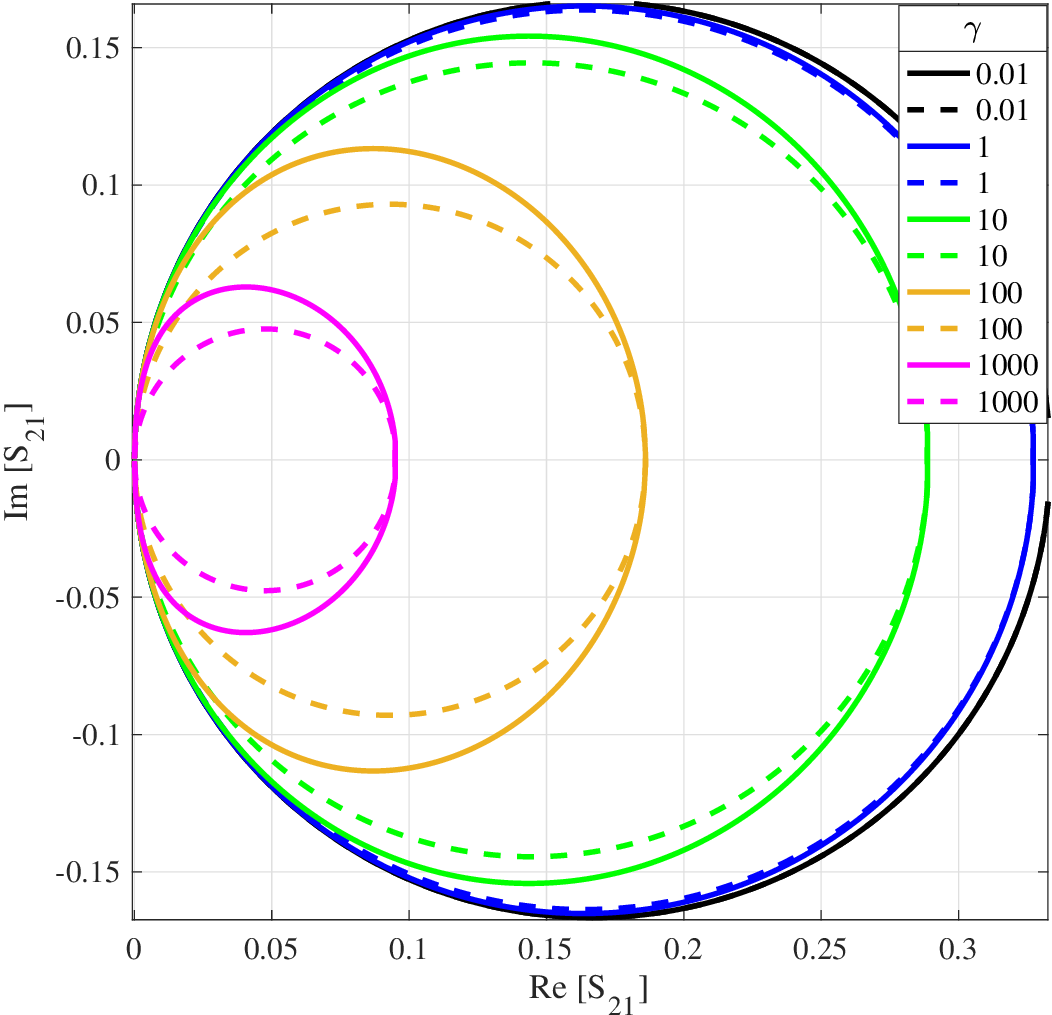}};
%
    	
 	 \end{tikzpicture}	

	\caption[]{\label{fig:S2_simulation_argand_versus_readout_power} $S_{21}$ in Argand complex plane as a function of frequency at different applied readout powers.  Solid lines: full QPH model; dashed lines: linear resonance curves calculated with the same fixed $Q_r$ as in Fig.\ref{fig:S2_simulation_magnitude_phase_versus_readout_power}.}
	
\end{figure}
Fig.\ref{fig:S2_simulation_argand_versus_readout_power} shows the resonance trajectories in the complex plane. At low power, the $S_{21}$ locus traces a nearly perfect circle. As the readout power increases, a distinctive oval‑shaped distortion appears. This distortion arises because the quasiparticle density (and hence the dissipation) increases as the frequency approaches resonance, reducing the radius of the trajectory. The trajectory starts on a larger‑radius circle, smoothly transitions onto a smaller‑radius circle near resonance, and then returns to the larger-radius off‑resonance circle. Therefore, the oval distortion observed in Fig.\ref{fig:S2_simulation_argand_versus_readout_power} provides a direct experimental signature of the QPH‑dominated nonlinear regime in superconducting resonators.

\section{Experimental measurement}\label{sec_3_measurement}
\begin{figure}[htbp!]
	\centering
    \begin{tikzpicture}
    	\node[inner sep=0, xshift=0cm, yshift=0cm] (image) at (0,0) {\includegraphics[width=.7\textwidth]{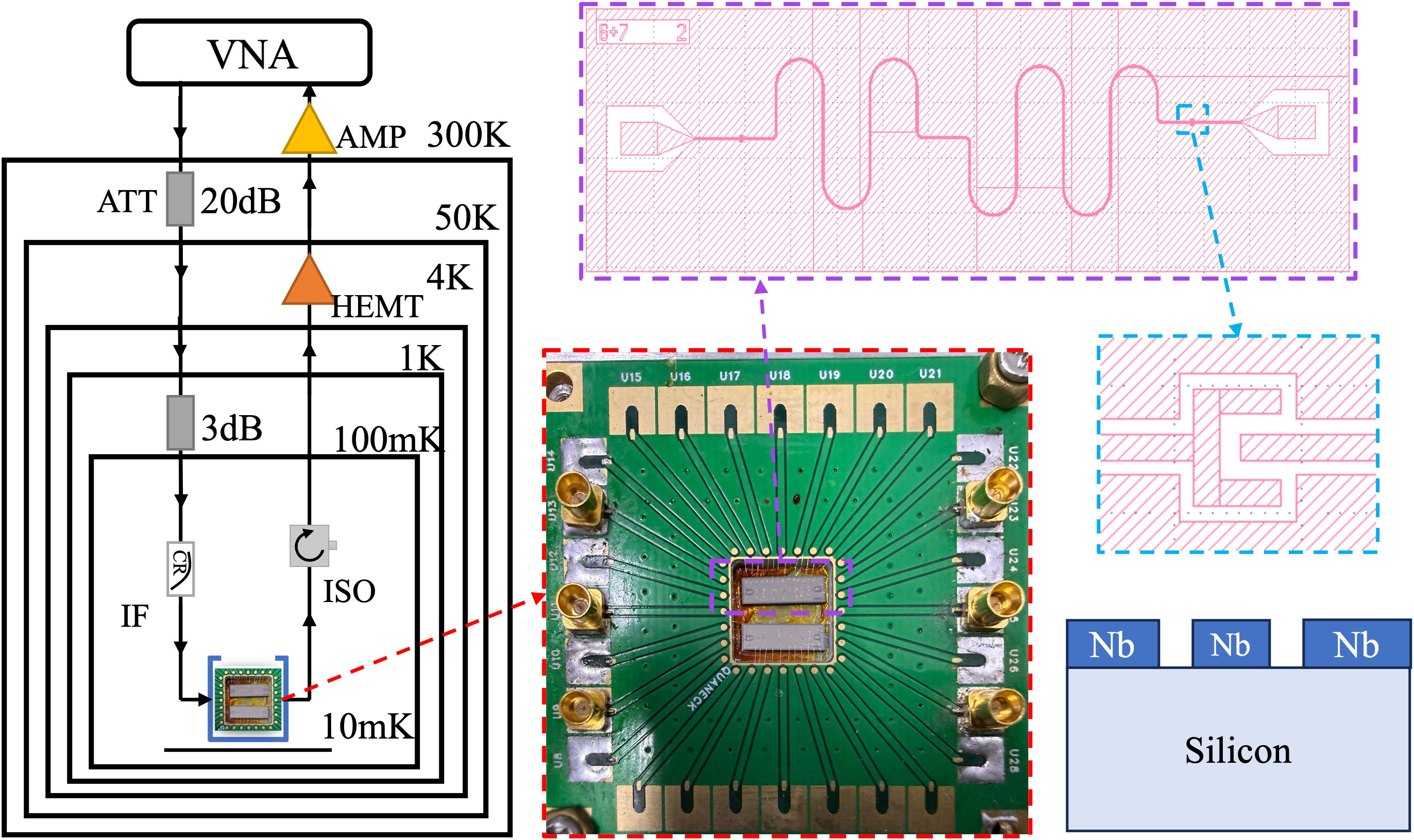}};
    \node[text=black, scale=1, align=left] at (-2.75, -3.5) {(a)};
    \node[text=black, scale=1, align=left] at (0.5, -3.5) {(b)};
    \node[text=black, scale=1, align=left] at (1.75, 0.8) {(c)};
    	 \node[text=black, scale=1, align=left] at (3.85, -1.2) {(d)};
    	  \node[text=black, scale=1, align=left] at (4, -3.5) {(e)};
    	  
 	 \end{tikzpicture}	

	\caption[]{\label{fig:S3_device_measurement_system} (a) Schematic diagram of the measurement system based on a dilution refrigerator. VNA: vector network analyser, ATT: attenuator, IR: infrared filter, ISO: isolator, HEMT: high electron mobility transistor, AMP: room‑temperature amplifier. (b) Photograph of the resonator chip wire‑bonded to the PCB. (c) Schematic of the CPW resonator mask layout. (d) Plan‑view of the stub coupling. (e) Cross‑section of the CPW line.}
\end{figure}
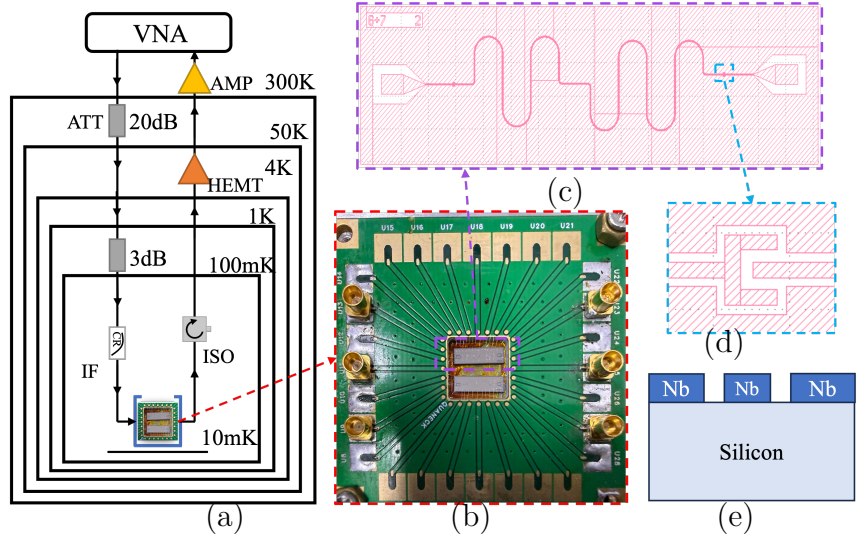

A CPW resonator was designed and fabricated on a high‑resistivity silicon substrate with a thickness of 500$\mathrm{\mu}$m. A 100nm‑thick niobium film (critical temperature $T_c \sim 8$K) was deposited and subsequently patterned by UV lithography and dry etching. As shown in the mask layout (Fig.\ref{fig:S3_device_measurement_system}(c)), the CPW design consists of a centre strip width of 10$\mathrm{\mu}$m and a gap width of 6$\mathrm{\mu}$m, and the total conductor length is $L = $ 13mm. The stub length for coupling control is 16$\mathrm{\mu}$m (Fig.\ref{fig:S3_device_measurement_system}(d)), and the cross‑sectional geometry of the CPW line is illustrated in Fig.\ref{fig:S3_device_measurement_system}(e). After dicing, the chip was mounted on a printed circuit board (PCB) and wire‑bonded using aluminium wires (Fig.\ref{fig:S3_device_measurement_system}(b)). The effective permittivity of the structure is estimated to be $\epsilon_{eff} \sim 6.3$, yielding a fundamental resonance frequency $f_r = c / 2 L \sqrt{\epsilon_{eff}} \sim$ 4.597GHz,  where $c$ is the speed of light in vacuum.

\begin{figure}[htbp!]
	\centering
    \begin{tikzpicture}
    	\node[inner sep=0, xshift=0cm, yshift=0cm] (image) at (0,0) {\includegraphics[width=.45\textwidth]{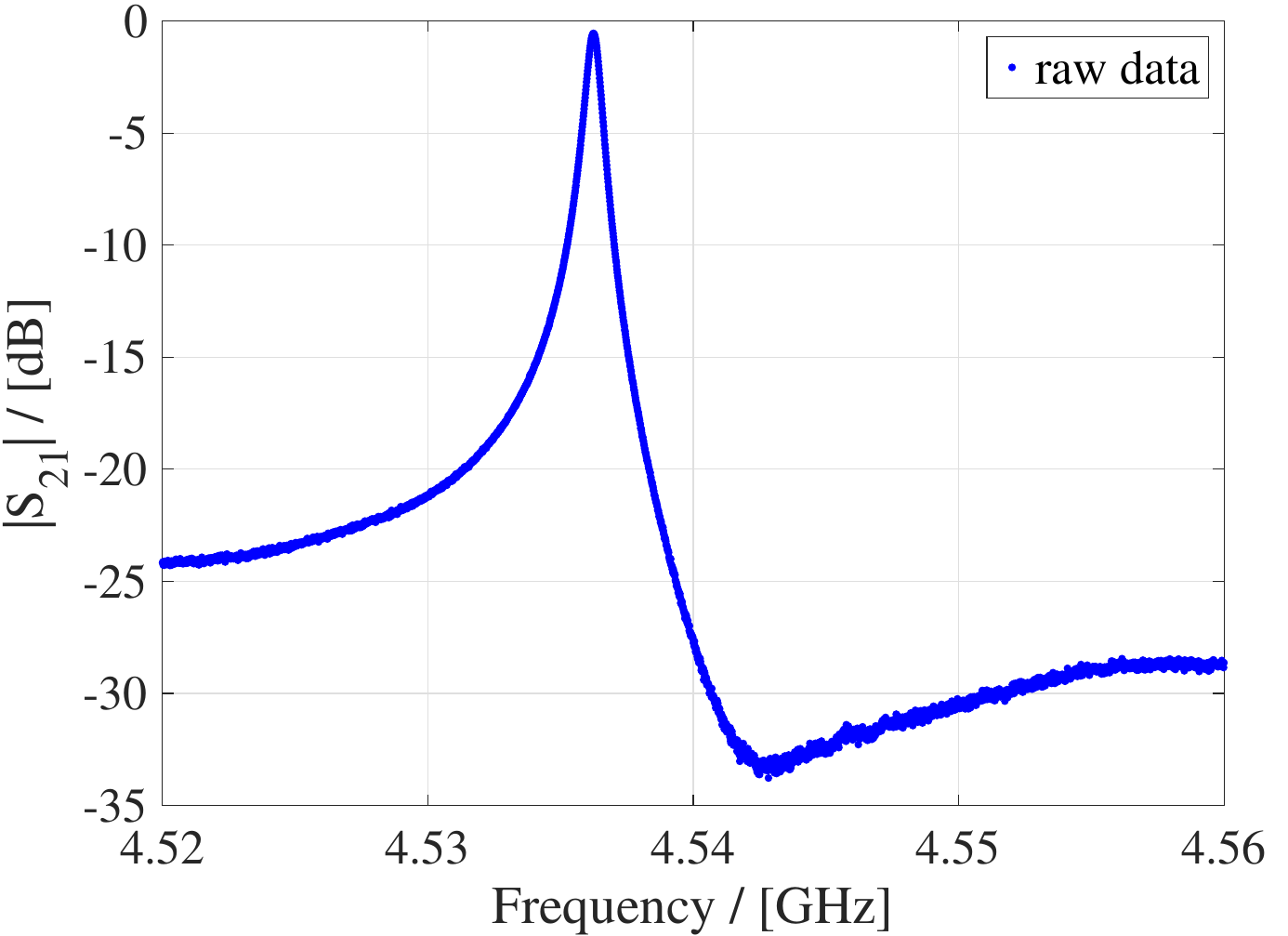}};
    	 \node[inner sep=0, xshift=8cm, yshift=0cm] (image) at (0,0) {\includegraphics[width= .45\textwidth]{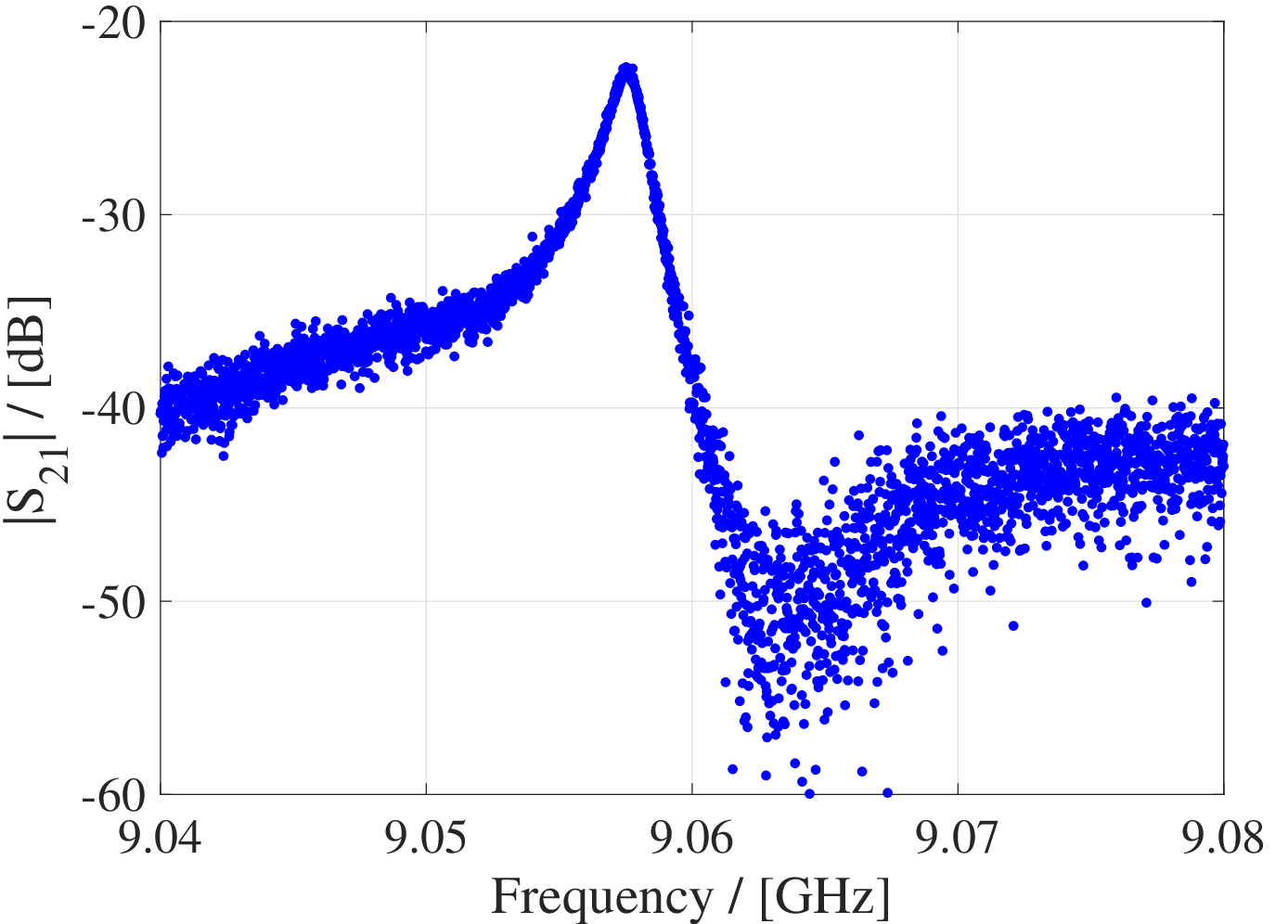}};
    	\node[text=black, scale=1, align=left] at (0, -2.75) {(a)};
    	\node[text=black, scale=1, align=left] at (8, -2.75) {(b)};
    	
 	 \end{tikzpicture}	

	\caption[]{\label{fig:S3_S21_first-order_and_second-order} Measured $S_{21}$ of the resonator at 20mK.  (a)Fundamental resonance. (b)Second-order harmonic.}
	
\end{figure}

The packaged resonator was cooled to a base temperature of 20mK in a dilution refrigerator. The measurement setup (Fig.\ref{fig:S3_device_measurement_system}(a)) comprises a vector network analyser (VNA) that supplies the readout signal; the signal passes through attenuators (ATT), infrared filters (IR) before reaching the resonator chip. The transmitted signal is amplified by a cryogenic high‑electron‑mobility transistor (HEMT) and a subsequent room‑temperature amplifier (AMP), and then measured by the VNA. Fig.\ref{fig:S3_S21_first-order_and_second-order} shows the fundamental resonance at 4.5363GHz, in close agreement with the design target and a second‑order harmonic is also observed at 9.0571GHz.

\begin{figure}[htbp!]
	\centering
    \begin{tikzpicture}
    	\node[inner sep=0, xshift=0cm, yshift=0cm] (image) at (0,0) {\includegraphics[width=.45\textwidth]{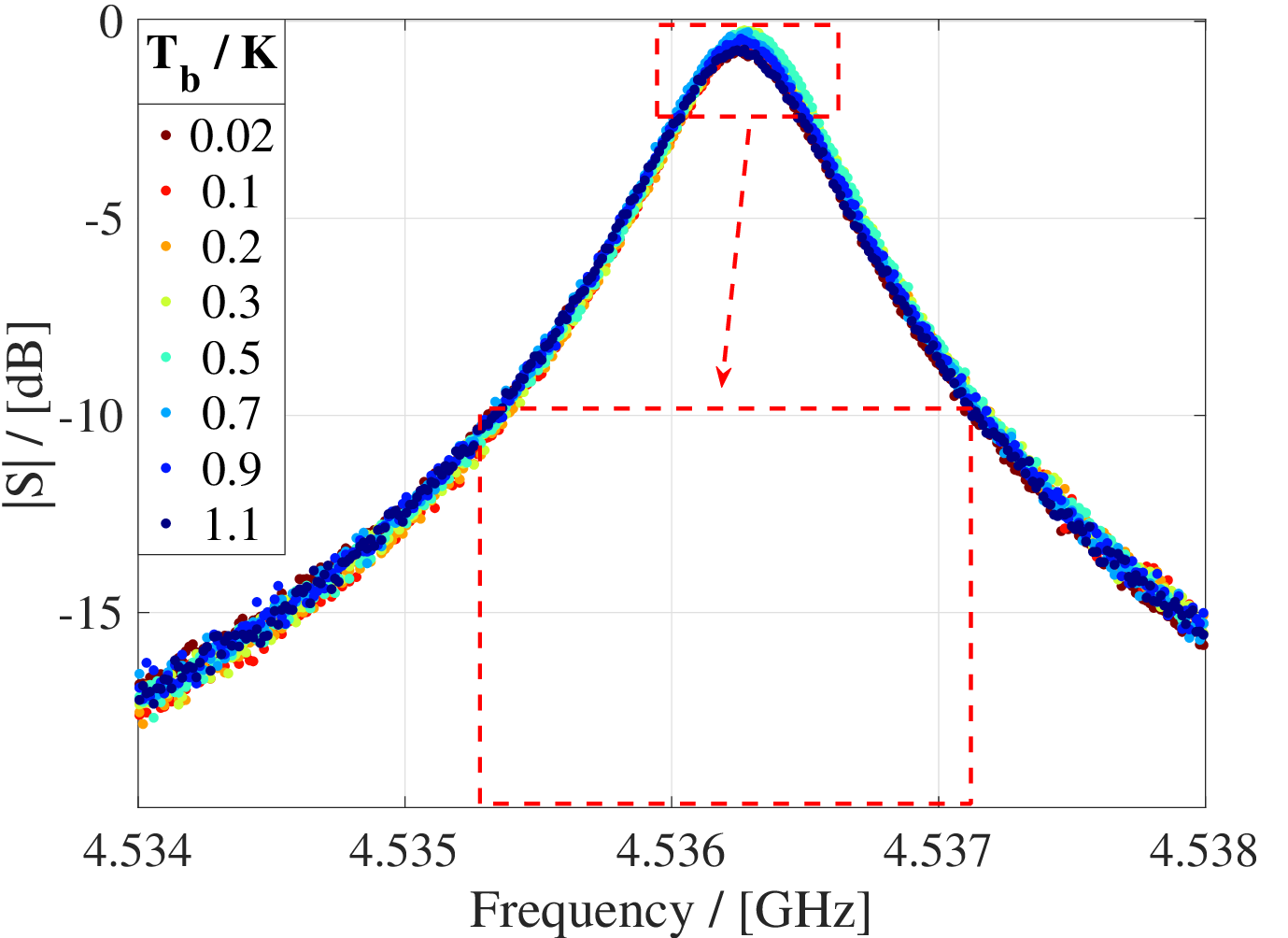}};
    	 \node[inner sep=0, xshift=0.5cm, yshift=-0.85cm] (image) at (0,0) {\includegraphics[width=.17\textwidth]{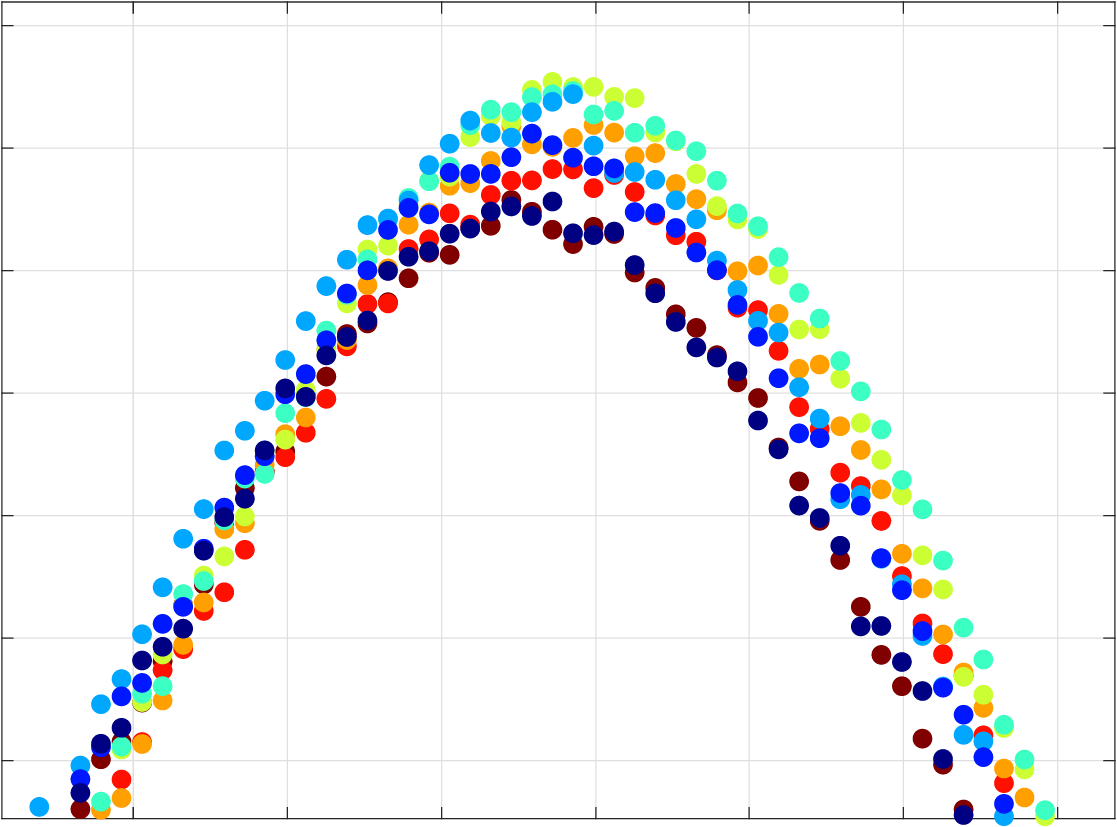}};
    	 \node[inner sep=0, xshift=8cm, yshift=0cm] (image) at (0,0) {\includegraphics[width= .45\textwidth]{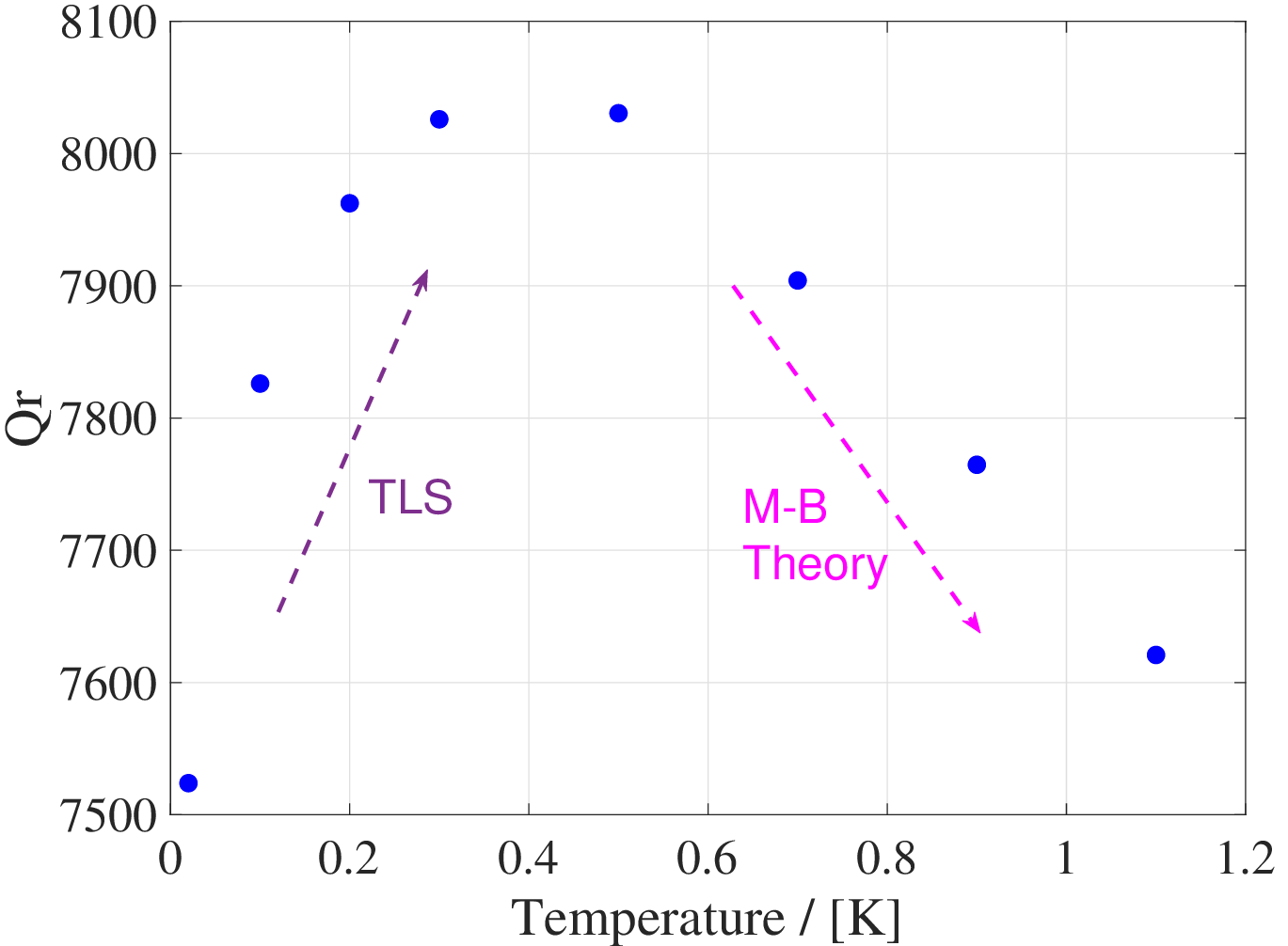}};
    	\node[text=black, scale=1, align=left] at (0, -2.75) {(a)};
    	\node[text=black, scale=1, align=left] at (8, -2.75) {(b)};
    	
 	 \end{tikzpicture}	

	\caption[]{\label{fig:S3_bath_temperature_sweep_measurement} (a) Measured transmission magnitude $S_{21}$ versus bath temperature $T_b$ at a fixed readout power of -70dBm. (b) Corresponding extracted quality factor $Q_r$ as a function of $T_b$.}
	
\end{figure}

Fig.\ref{fig:S3_bath_temperature_sweep_measurement} presents the temperature dependence of the resonance response and the extracted quality factor for the half‑wavelength resonator at a readout power of -70dBm. At base temperature (20mK), the resonance is centred at 4.5363GHz with an initial quality factor $Q_r \sim$ 7500. As shown in Fig.\ref{fig:S3_bath_temperature_sweep_measurement}(a), the resonance peak amplitude first increases with rising bath temperature, reaching a maximum in the range of 300-500mK, and then decreases at higher temperatures. Correspondingly, Fig.\ref{fig:S3_bath_temperature_sweep_measurement}(b) shows that the extracted quality factor initially rises at low temperatures, consistent with the saturation of two‑level system losses. With further increase in $T_b$, the quality factor decreases, indicating that thermally excited non-equilibrium quasiparticle losses become increasingly dominant. This behaviour is in qualitative agreement with the Mattis–Bardeen theory, which predicts an exponential increase in the quasiparticle density and hence the surface resistance with rising temperature \cite{mattis1958theory}. Similar temperature-dependent degradation of the quality factor has been observed in various superconducting resonator systems \cite{fischer2024nonequilibrium, sun2026quasiparticle, li2025low, alexander2025power}.
\begin{figure}[htbp!]
	\centering
    \begin{tikzpicture}
    	\node[inner sep=0, xshift=0cm, yshift=0cm] (image) at (0,0) {\includegraphics[width=.45\textwidth]{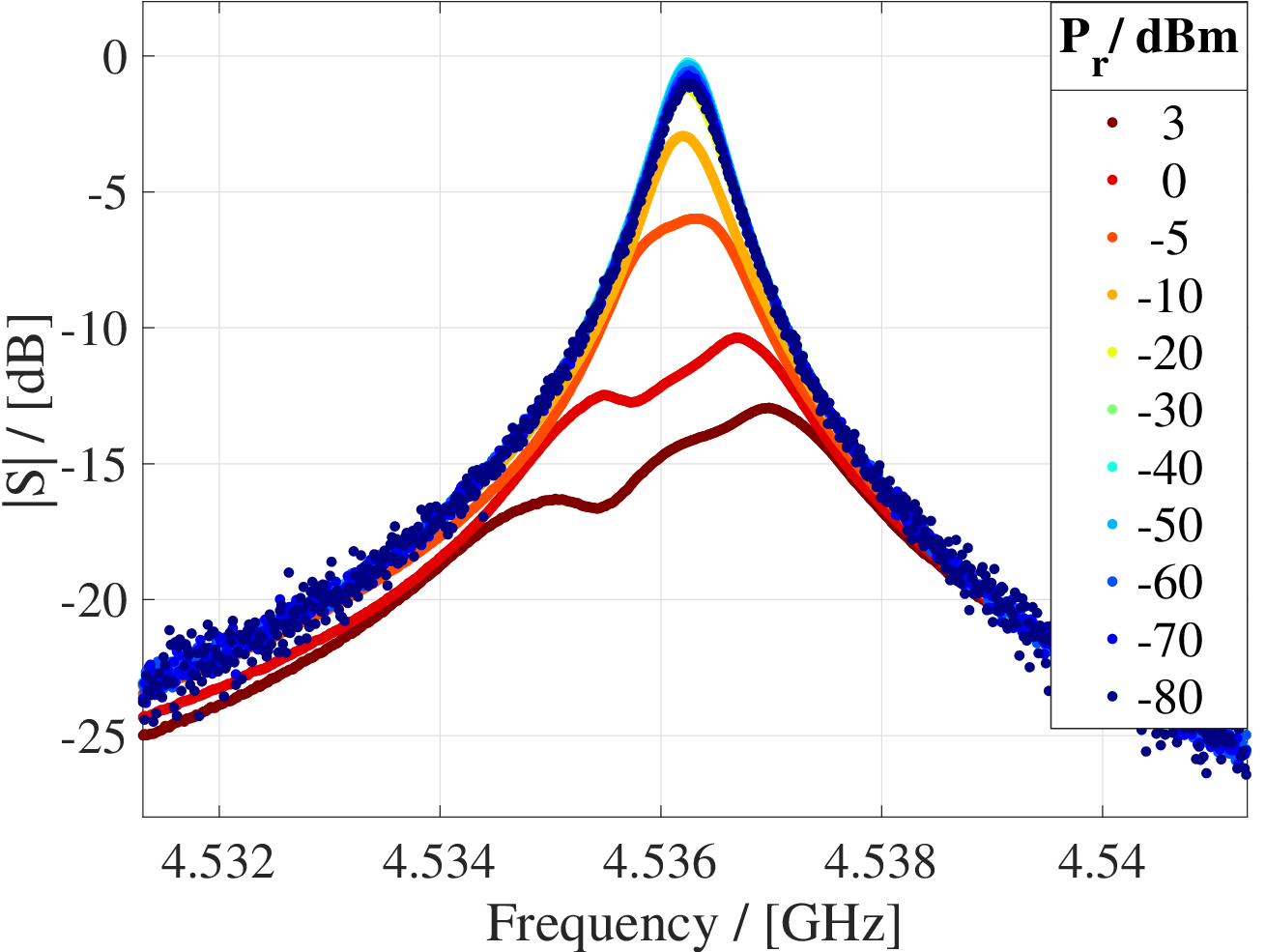}};
    	 \node[inner sep=0, xshift=8cm, yshift=0cm] (image) at (0,0) {\includegraphics[width= .44\textwidth]{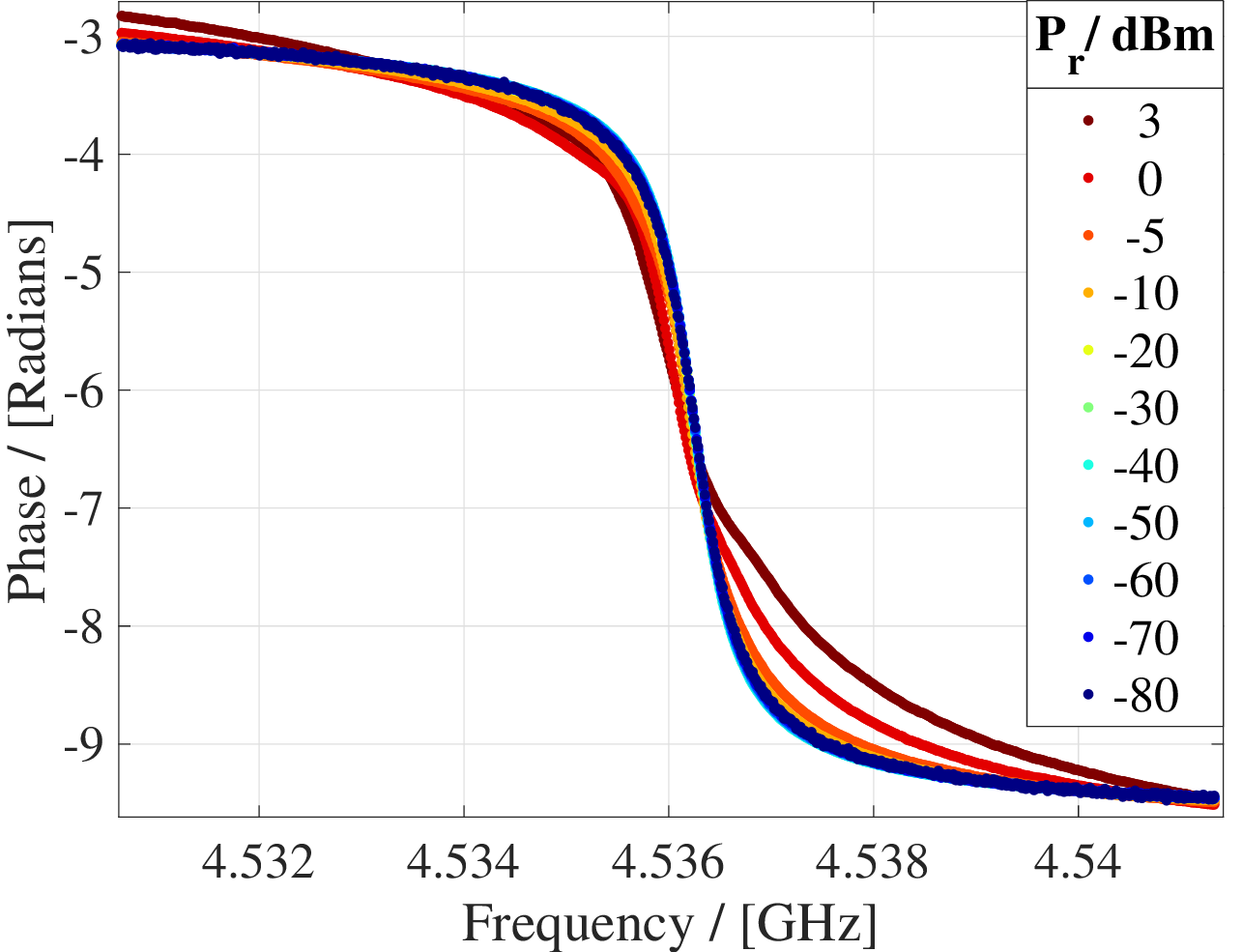}};
    	\node[text=black, scale=1, align=left] at (0, -2.85) {(a)};
    	\node[text=black, scale=1, align=left] at (8, -2.85) {(b)};
    	
 	 \end{tikzpicture}	

	\caption[]{\label{fig:S3_readout_power_sweep_measurement_magnitude_and_phase} Measured transmission magnitude $S_{21}$ (a) and phase (b) versus readout power at 20mK.}
	
\end{figure}
Fig.\ref{fig:S3_readout_power_sweep_measurement_magnitude_and_phase} presents the experimental resonance response as a function of readout power at base temperature (20 mK), where TLS losses are still present at low readout powers. In Fig.\ref{fig:S3_readout_power_sweep_measurement_magnitude_and_phase}(a), the resonance magnitude initially increases with increasing power, consistent with TLS saturation. As the power continues to rise, the magnitude decreases sharply, indicating that quasiparticle heating losses become dominant. At even higher powers, the resonance lineshape becomes distorted. Correspondingly, Fig.\ref{fig:S3_readout_power_sweep_measurement_magnitude_and_phase}(b) shows that the phase response also exhibits obvious distortion under high-power excitation. These experimental observations are in good qualitative agreement with the theoretical predictions shown in Fig.\ref{fig:S2_simulation_magnitude_phase_versus_readout_power} confirming that the device transitions from a TLS-limited regime at low power to a QPH-dominated regime at high readout powers, where the nonlinear effects induced by microwave-induced quasiparticle generation become manifest.

\begin{figure}[htbp!]
	\centering
    \begin{tikzpicture}
  	    	 \node[inner sep=0, xshift=0cm, yshift=0cm] (image) at (0,0) {\includegraphics[width= .45\textwidth]{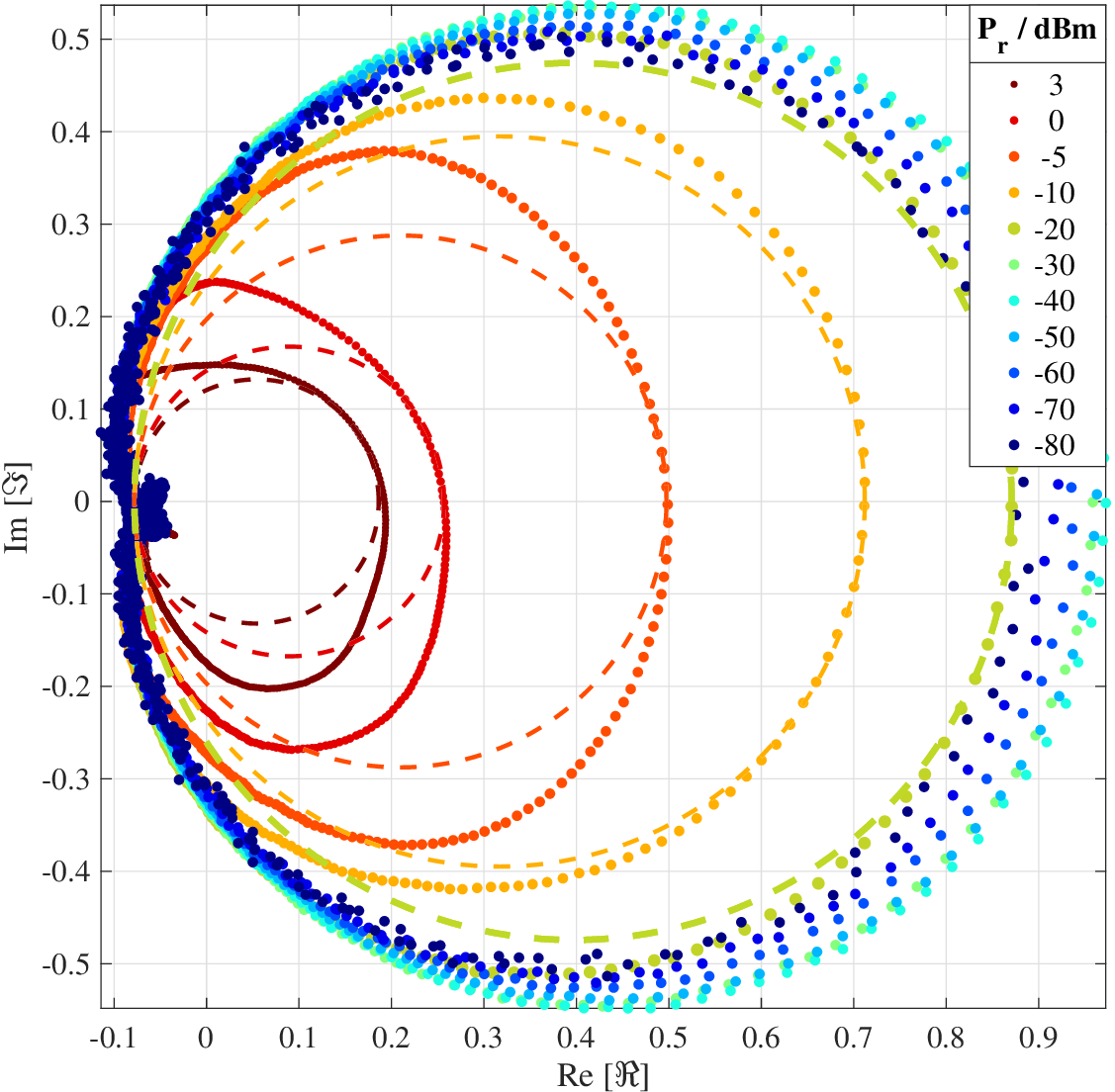}};
  	    	 
    	\node[inner sep=0, xshift=8cm, yshift=-0cm] (image) at (0,0) {\includegraphics[width=.45\textwidth]{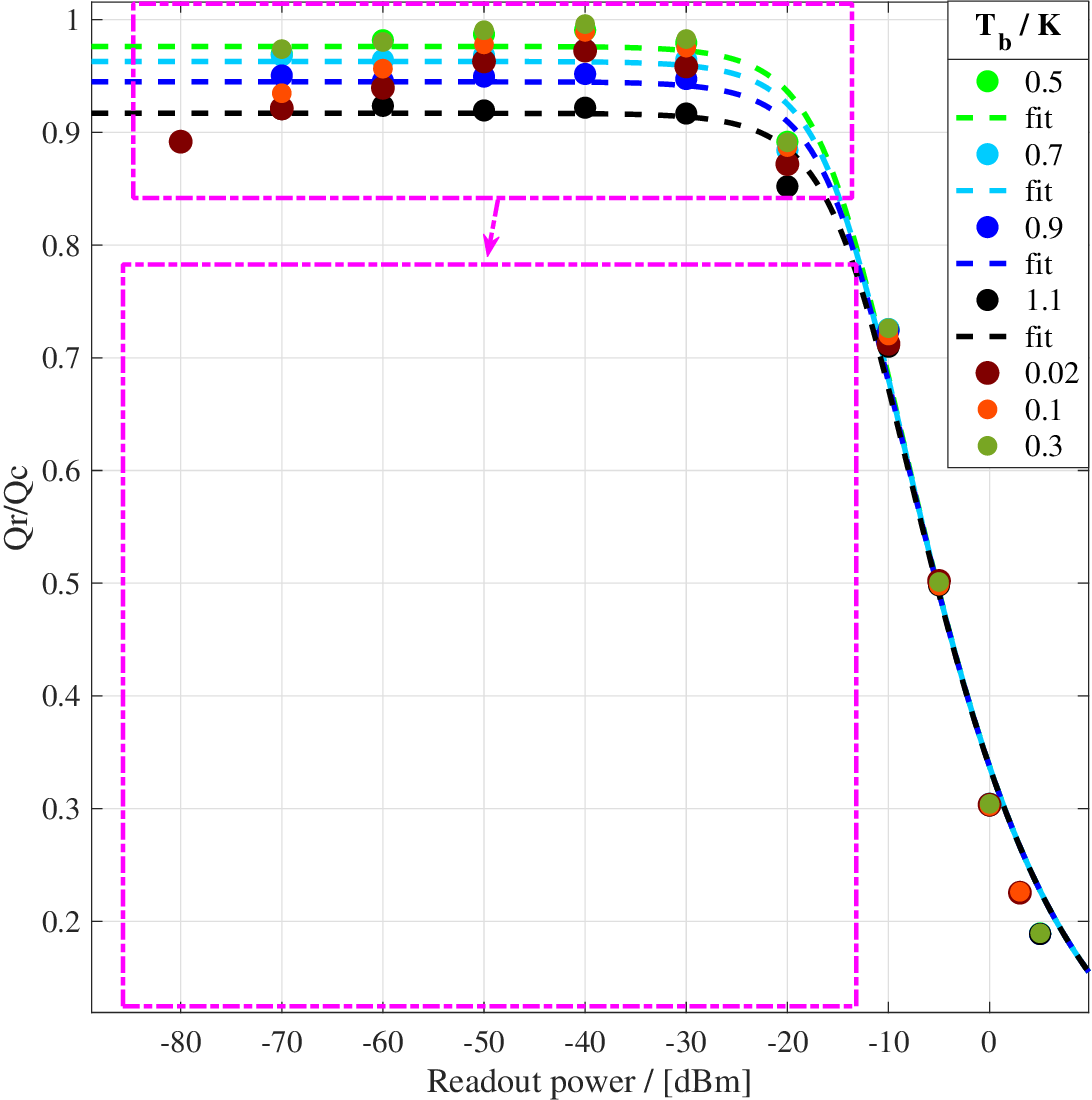}};
    	 \node[inner sep=0, xshift=7.65cm, yshift=-0.55cm] (image) at (0,0) {\includegraphics[width=.295\textwidth]{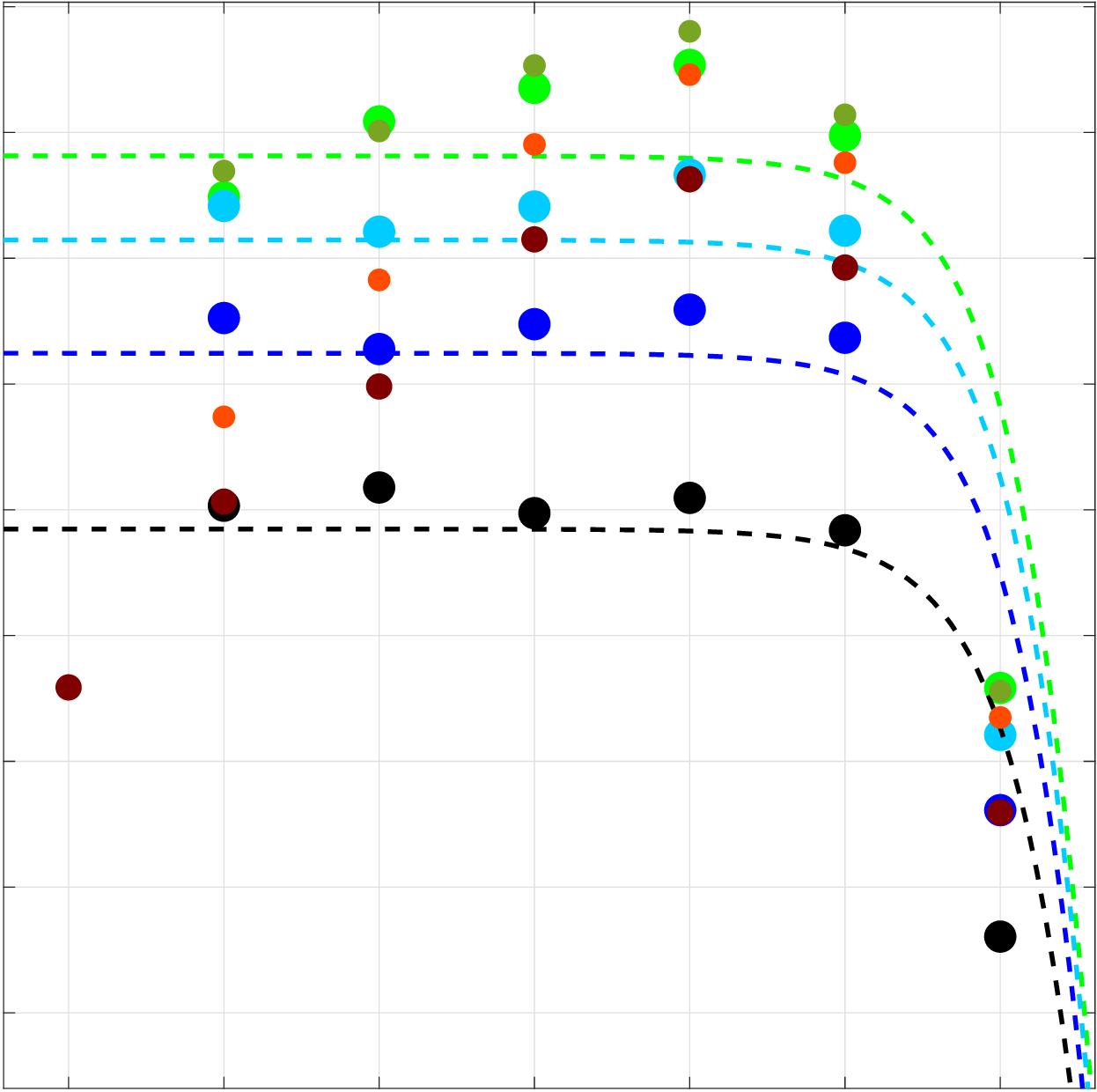}};

    	\node[text=black, scale=1, align=left] at (0, -3.75) {(a)};
    	\node[text=black, scale=1, align=left] at (8, -3.75) {(b)};
    	
 	 \end{tikzpicture}	

	\caption[]{\label{fig:S3_readout_power_sweep_measurement_Argand_Extracted_Qfactor} Measured $S_{21}$ trajectories in the Argand complex plane at 20mK for varying readout power. (b) Extracted quality factors as a function of readout power at different bath temperatures. The experimentally extracted Q factors are shown as dot markers, while the fits to the proposed model are represented by dashed lines.}
	
\end{figure}

Fig.\ref{fig:S3_readout_power_sweep_measurement_Argand_Extracted_Qfactor}(a) presents the measured $S_{21}$ resonance trajectories in the Argand complex plane at 20mK for varying readout power. In the low-power regime, the diameter of the resonance circle increases with increasing power, corresponding to an increase in the total quality factor as two-level system losses are gradually saturated by the readout power (reaching saturation around -40dBm).  As the power increases further, the circle diameter decreases and the trajectory progressively develops into an oval-shaped distortion. At even higher readout powers (exceeding 0dBm), the resonance distortion becomes more pronounced; while the overall morphology remains qualitatively consistent with the QPH model, additional mechanisms—such as kinetic inductance nonlinearity, vortex dynamics, or weak links—may contribute to the observed behaviour. A detailed investigation of these high-power effects is beyond the scope of the present work but represents an important direction for future study.

Fig.\ref{fig:S3_readout_power_sweep_measurement_Argand_Extracted_Qfactor}(b) shows the extracted quality factor as a function of readout power at different bath temperatures. In the low-temperature regime, increasing the bath temperature helps saturate TLS losses, thereby raising the power level at which quasiparticle heating losses begin to dominate. As the bath temperature increases further, the thermally excited quasiparticles become the dominant loss mechanism after TLS saturation, and the crossover point shifts to lower quality factor levels. Once QPH losses take over, all curves collapse onto a common decreasing trend with increasing readout power. The observed oval-shaped distortion in the Argand plane, together with the systematic evolution of the quality factor with readout power at different bath temperatures, is in good agreement with the predictions of our theoretical model across the parameter range where QPH is expected to be the dominant dissipative mechanism.

\begin{table}[htbp]
\centering
\caption[Best-fit values for $Q_{qp,th}/Q_c$]{Best-fit values of $Q_{qp, th} / Q_c$ for readout power sweep data at different bath temperatures above 500\,mK, where a clean measurement of the QPH-limited quality factor regime is obtained. Additional inputs to the model include the power scaling parameter $P_c = 0.0311$\,mW and $Q_{\text{other}} / Q_c = 1.1$, both obtained from a single global fit and held fixed across all bath temperatures and readout powers.}
\label{table_fitting_parameter}
\begin{tabular}{lcccc}
\toprule
$T_b$ / [K] & 0.5 & 0.7 & 0.9 & 1.1 \\
\midrule
$Q_{qp, th} / Q_c$ & 8.6791 & 7.7265 & 6.7009 & 5.5095 \\
\bottomrule
\end{tabular}
\end{table}
The parameters used in the fitting procedure are summarised in Table.\ref{table_fitting_parameter} and its caption. Above 500mK, TLS effects are fully saturated, yielding a clean measurement of the QPH-limited quality factor regime, where we fit the extracted Q factor versus readout power at different bath temperatures. The global parameters $P_c = $0.0311mW and $Q_{other} / Q_c = $1.1 were obtained from a single global fit and held fixed across all bath temperatures and readout powers. Within this framework, $Q_{qp,th} / Q_c$ was the only free parameter allowed to vary with temperature, and it was determined by best fit to the measured data at each bath temperature. This parameter represents the contribution of thermally excited quasiparticle loss mechanisms to the measured resonator response. The well agreement between the fits and the experimental data further demonstrates the consistency between our theoretical model and the experimental results.

\section{Conclusion}
In this work, we have validated a macroscopic model based on the modified Rothwarf-Taylor equations to describe the dissipative non-linearity in a superconducting Nb half-wavelength resonator due to quasiparticle heating. The model establishes a direct quantitative link between the resonator quality factor and the microwave readout power,  by explicitly tracking the power-dependent quasiparticle generation rate. The key outcome of our study is the identification and experimental confirmation of an oval-shaped distortion in the resonance circle within the Argand complex plane. We demonstrate that this characteristic distortion serves as a practical experimental signature of the QPH-dominated nonlinear regime, providing a useful means for diagnosing the onset of power-induced losses in similar devices. While additional nonlinear mechanisms may become relevant at the very highest readout powers, the QPH model captures the dominant dissipative behaviour over a wide range of operating conditions relevant to most practical applications.

Our experimental measurements, conducted on a fabricated Nb CPW resonator over a wide range of bath temperatures (20 mK to 1.1 K) and readout powers, corroborate the theoretical predictions. At low temperatures, the quality factor initially increases with readout power due to the saturation of TLS losses. As the power is further increased, we observe a sharp decrease in Q factor and the emergence of the distinctive oval distortion in the $S_{21}$ trajectory, in good agreement with our QPH model. The model successfully fits the extracted quality factors across different bath temperatures and readout powers, indicating that QPH is the dominant loss mechanism once TLS effects are saturated, consistent with the morphological signatures identified in the Argand plane. 

The well agreement between theory and experiment across a wide parameter space demonstrates the predictive power of our framework. Our findings not only provide a comprehensive understanding of the dissipative non-linearity induced by QPH but also offer a practical experimental methodology for identifying the onset of this regime. This is particularly valuable for optimizing the operating point of superconducting devices, such as KIDs, where balancing high readout power for noise suppression against QPH-induced losses is crucial for achieving peak performance. Future work will extend this framework to include other loss mechanisms, such as kinetic inductance non-linearity and vortex dynamics, which may become relevant at even higher readout powers, and will explore the temporal dynamics of these systems under pulsed operation.

\section{Acknowledgements}
The authors thank Dr. Thomas and Prof.Withington for proposing concept of the theoretical framework and their guidance to Dr.Z Sun throughout the theoretical and experimental investigations. 

The authors also thank Chengdu Data Automation System Technologies Go.Ltd for providing cryogenic amplifier for experimental measurements.

\bibliographystyle{iopart-num}
\bibliography{reference}


\end{document}